\documentclass{article}

\usepackage{PRIMEarxiv}

\usepackage[utf8]{inputenc} % allow utf-8 input
\usepackage[T1]{fontenc}    % use 8-bit T1 fonts
\usepackage{hyperref}       % hyperlinks
\usepackage{url}            % simple URL typesetting
\usepackage{booktabs}       % professional-quality tables
\usepackage{amsfonts}  
\usepackage{array} % blackboard math symbols
\usepackage{nicefrac}       % compact symbols for 1/2, etc.
\usepackage{microtype}      % microtypography
\usepackage{lipsum}
\usepackage{fancyhdr}       % header
\usepackage{graphicx}       % graphics
\graphicspath{{media/}}     % organize your images and other figures under media/ folder
\usepackage{multicol}
\title{Advancing Dermatological Diagnosis: Development of a Hyperspectral Dermatoscope for Enhanced Skin Imaging
}

\author{
  Martin J. Hetz  \\
  Digital Biomarkers for Oncology Group  \\
  German Cancer Research Center (DKFZ) \\
  Heidelberg, Germany \\
  \texttt{{martinjoachim.hetz}@dkfz-heidelberg.de} \\
  \AND
  Carina  Nogueira Garcia \\
  Digital Biomarkers for Oncology Group \\
  German Cancer Research Center (DKFZ) \\
  Heidelberg, Germany \\
  \texttt{carina.nogueiragarcia@dkfz-heidelberg.de} \\
  \AND
  Sarah Haggenm\"uller \\
  Digital Biomarkers for Oncology Group \\
  German Cancer Research Center (DKFZ) \\
  Heidelberg, Germany \\
  \texttt{sarah.haggenmueller@dkfz-heidelberg.de} \\
  \AND
  Titus J. Brinker \\
  Digital Biomarkers for Oncology Group \\
  German Cancer Research Center (DKFZ) \\
  Heidelberg, Germany \\
  \texttt{titus.brinker@nct-heidelberg.de} \\
}

\begin{document}
\maketitle

\begin{abstract}
Clinical dermatology necessitates precision and innovation for efficient diagnosis and treatment of various skin conditions. This paper introduces the development of a cutting-edge hyperspectral dermatoscope (the Hyperscope) tailored for human skin analysis. We detail the requirements to such a device and the design considerations, from optical configurations to sensor selection, necessary to capture a wide spectral range with high fidelity. Preliminary results from 15 individuals and 160 recorded skin images demonstrate the potential of the Hyperscope in identifying and characterizing various skin conditions, offering a promising avenue for non-invasive skin evaluation and a platform for future research in dermatology-related hyperspectral imaging.
\end{abstract}

% keywords can be removed

\section{Introduction}
Skin cancer is one of the most prevalent cancer types worldwide, mostly consisting of melanoma and non-melanoma (NMSC) forms \cite{SkinCancer}. As its incidence rates are on the rise, so is the need for early detection and treatment, which are highly dependent on efficient diagnostic techniques \cite{Urban2021-ty, Tripp2016-am}. While less frequent than NMSC, such as squamous cell carcinoma and basal cell carcinoma, melanoma has a higher death rate if not diagnosed early \cite{SkinCancer, American_Academy_of_Dermatology_Ad_Hoc_Task_Force_for_the_ABCDEs_of_Melanoma2015-tw}. Conventional diagnosis techniques, such as dermatoscopy and visual examination, mostly depend on the physician’s knowledge and training \cite{Kittler2002-vb}. However, this is not a problem of lack of training, as sensitivity levels above 80\% are rarely achieved even by experienced dermatologists \cite{Vestergaard2008-yl}.  Early and precise detection - already on a clinical level - is the key to improving the prognosis for melanoma patients \cite{SkinCancer2}. \\ \newline
In cases where a clinical suspicion of malignancy cannot be ruled out, histopathological examination is performed, serving as the gold standard for the diagnosis of melanoma and other types of skin cancer \cite{Mayer1997-ke, Vestergaard2008-yl}. However, this requires the excision of the respective suspicious lesion which often causes stress and discomfort for the patient~\cite{Augustin1999-bp, Vaidya2019-bk}. Non-invasive diagnostic procedures may be useful in minimizing patient discomfort while enabling efficient examinations. \\ \newline
Therefore, there is a pressing need for new non-invasive techniques capable of detecting skin cancer at its earliest stages and with a high degree of accuracy. To address this problem, several promising techniques have already been presented \cite{Heibel2020-ng, Soglia2022-vf}. One of these promising techniques is hyperspectral imaging (HSI), which simultaneously allows a spatially and spectrally resolved measurement of incident light. HSI cameras combine the spatially resolved measurement of reflected light, such as a RGB camera, and the spectrally resolved measurement of an optical spectrometer, which are used for analyzing the composition of light. A HSI camera records a spectrum for each pixel on a 2D sensor array, resulting in a 3D data cube called a hypercube. Analyzing the reflectance of each pixel allows drawing conclusions about the properties of the captured tissue \cite{Lu2014-wp}. 
\section{Related work}
\label{sec:background}
HSI is playing an increasingly important role in medicine and already has several medical applications \cite{Lu2014-wp, Halicek2019-zl, Yoon2022-ea}. Due to the easy accessibility and the in-vivo character of optical measurements of the skin, the optical diagnosis of skin diseases is an obvious choice \cite{Aggarwal2022-ji}. In the recent past, multiple research projects have been carried out in the field of HSI for skin lesions. The work is very wide-ranging, from the differentiation of benign and malignant skin lesions~\cite{Lindholm2022-az, Calin2022-im, Torti2020-oa, Courtenay2021-uh, Hosking2019-ua}, the classification of skin lesions \cite{Huang2024-pj, Huang2023-uq, Rasanen2021-px, Leon2020-tr}, to the estimation of skin chromophores and other biological parameters \cite{Gevaux2021-zm, Zherebtsov2019-ds}. The work differs not only in the application, but also greatly in the devices used. The devices vary in terms of image acquisition methods, spatial and spectral resolution or whether the devices are handheld or stationary \cite{Lu2014-wp}. The basic structure of a setup for recording HSI consists of a camera, optics, a dispersive element, lighting and a computer connected to the camera \cite{Fabelo2019-ug, Hosking2019-ua, Vasefi2014-kp, Zherdeva2016-eb, Aloupogianni2021-pk, Raita-Hakola2022-he}. In addition to the use of dedicated hardware, some works combine RGB images with optical models to derive spectral data from the RGB data. This can be achieved by means of standardized spectra \cite{Huang2023-uq} as well as calibration charts \cite{Huang2024-pj}. Lindholm et al. apply a device that is based on a spectral scanning method and covers a wavelength range from 475nm to 975nm \cite{Raita-Hakola2022-he}. They use pixel-by-pixel analysis to determine whether the lesion is benign or malignant, both for pigmented and non-pigmented lesions \cite{Lindholm2022-az}. Huang et al. utilise a digital camera, a spectrograph and a calibration chart to derive spectral information from the RGB images. This data is then used to distinguish mycosis fungoides from psoriasis and atopic dermatitis \cite{Huang2024-pj}. Aloupogianni et al. use a custom acquisition system with a spectral range of 420 nm to 750 nm to acquire images of lesions directly after excision. The acquired data is used to develop a tumor margin detection \cite{Aloupogianni2022-ah}. Calin et al. employ a spatially scanning, also known as pushbroom, HSI system that covers the wavelength range from 400 nm to 800 nm. They then attempt to differentiate between healthy skin and BCC using unsupervised learning \cite{Calin2022-im}. Torti et al. use a hand-held system based on the snapshot method from Fabelo et al. \cite{Fabelo2019-ug} with a covered wavelength range of 450 nm to 950 nm and a spatial resolution of $50\times 50$ pixels. The recorded pigmented skin lesions are then differentiated into benign and malignant or atypical \cite{Torti2020-oa}. The works by Hosking et al. presents the Melanoma Advanced Imaging Dermatoscope, which is used to classify pigmented skin lesions. The system records 21 channels in the 350 nm to 950 nm range using a spectral scanning principle \cite{Hosking2019-ua}.  Leon et al. use the imager from Fabelo et al. \cite{Fabelo2019-ug} to automatically differentiate pigmented skin lesions into benign or malignant. R\"as\"anen et al. use a spectral scanning prototype which records 76 bands in the range 450 nm to 850 nm. In this work, a distinction was made between pigmented BCC and melanoma \cite{Rasanen2021-px}. In addition to the classification of skin lesions, the estimation of skin parameters plays a major role. Gevaux et al. use a spectral scanning system that captures high-resolution images in the range between 400 nm and 700 nm. By using an optical model of the skin, oxygen rate, blood volume fraction and melanin concentration are then estimated \cite{Gevaux2021-zm}. A similar approach is applied in the work of Zherebtsov et al., which deals with the estimation of skin chromophores and blood oxygenation. A system based on a snapshot hyperspectral camera with a range of 500 nm to 900 nm \cite{Zherebtsov2019-ds} is used. \\ \newline 
Researchers employ a range of devices with diverse properties, yet determining the optimal configuration remains challenging. Therefore, we have developed a device closely mirroring conventional dermatoscopes utilised in clinical settings. This device boasts an exceptional spectral range and resolution, enabling the exploitation of spectral features in comparison to alternative device.  With the Hyperscope, we achieve the highest spatial resolution among handheld devices which is specifically relevant for skin analysis, as morphological features of the skin are captured in greater detail at higher spatial resolution.
\section{Conceptualization}
\label{sec:conceptualization}
In order to develop a hyperspectral dermatoscope that can be used for various skin diseases, several aspects must be taken into consideration. These aspects include the optical setup as well as the selection of the imaging method or wavelength range used. In the following, we first introduce the fundamental concepts of these aspects, outline construction and design alternatives, and subsequently detail the implementation process tailored to our Hyperscope. By doing so, our goal is to introduce a hyperspectral dermatoscope that is seamlessly integrable into clinical practice while simultaneously fulfilling stringent technical requirements.  
\subsection{Measurement principle}
\label{sec:measurement_principle}
Manufacturers employ various measurement principles to acquire HSIs, with scan-based or snapshot-imaging-based methods being commonly used \cite{Yoon2022-ea, Lu2014-wp}. The different measurement principles result in several characteristics of the hypercube. \\ \newline
Scanning methods, on one hand, allow superior spatial resolution compared to snapshot hyperspectral cameras. This higher spatial resolution enables better detection of skin morphological features. Within the category of scanning methods, a further distinction is made between spatial scanning methods and spectral scanning methods.Spatial scanning methods capture the complete spectrum of pixels either  pointwise or linewise, while  spectral scanning methods sequentially record complete images at different wavelengths. Spectral scanning methods provide the advantage of displaying a preview image in order to align the image capturing system with the location to be recorded. However, both types of scanning methods may encounter problems for in-vivo measurements due to motion artifacts resulting from  high acquisition time \cite{Lu2014-wp}. \\ \newline 
Snapshot hyperspectral cameras, on the other hand, allow the entire hypercube to be captured in a single shot, thus enabling image acquisition in real-time. Their short exposure times make them ideal for hand-held devices, minimizing temporal artifacts caused by patients or user motion \cite{Johnson2007-ym, Hagen2013-zy}.  However, snapshot hyperspectral cameras may exhibit lower spatial and spectral resolution compared to scanning cameras \cite{Lu2014-wp}. \\ \newline
Given our focus on  a simple and portable imaging approach,  specifically designed for practical use in clinical routine, we opted for a hyperspectral camera based on the snapshot method. While this choice results in  lower spatial resolution compared to other recording methods, it is only of  subordinate importance  for our application, as we plan to analyse measurements using deep learning methods, typically operating at  spatial resolutions around $224\times 224$ range \cite{Dosovitskiy2020-gr, He2015-ba, Krizhevsky2012-wg}. Another essential advantage of the snapshot method is that no motion artefacts occur during image acquisition. Moreover, its immunity to motion artifacts  makes it easier for clinical staff to operate the Hyperscope, as minor movements during image acquisition do not comprimise  the sharpness of the images.
\subsection{Spatial resolution and field of view}
\label{sec:spatial_resolution}
Further essential aspects to consider are the spatial resolution and the field of view (FoV) of the hyperspectral dermatoscope. The sensor resolution must always be considered together with the FoV, as a relatively high spatial resolution can be achieved by combining a low sensor resolution with a small field of view. Given technical standards, it may be difficult to visualise more than three channels. Therefore, it is advisable to leverage algorithms to evaluate the HSI. Due to the high memory requirements, particularly for deep neural networks (DNNs), smaller image resolutions have become established. A frequently used resolution in this context is $224 \times 224$ pixels \cite{Dosovitskiy2020-gr, He2015-ba, Krizhevsky2012-wg}, which constitutes the lower limit for spatial sensor resolution. \\ \newline
Together with the FoV and the spatial resolution, the actual resolution of the structure size can be determined. When selecting the FoV, both the size of the lesions to be recorded and the maximum size of the contact area must be considered in order to be able to record lesions on curved surfaces, such as the nose. According to Heinlein et al. the diameter of the majority of cancerous skin lesions is below 1.5 cm \cite{Heinlein2024-mg}. The FoV should therefore approximately cover this size to provide a hyperspectral dermatoscope that can be used for a wide range of skin diseases. \\ \newline
Based on the requirements, we decided in favour of a camera with a resolution slightly exceeding the minimum mentioned. This simultaneously ensures an improved recording of morphological features of the skin. The exact spatial resolution of the camera is $290 \times 275$ pixels. For the FoV, we opted for a square FoV  of 2 cm × 2 cm, enabling the coverage of larger lesions (e.g. large congenital nevi) as well. The square FoV allows us to record images without borders. The selected FoV and the sensor resolution result in an area resolution of around 200 $pixels/mm^2$.
\subsection{Spectral properties}
\label{sec:spectral_properties}
\begin{figure*}[!t]
\centering
\includegraphics[width=\textwidth]{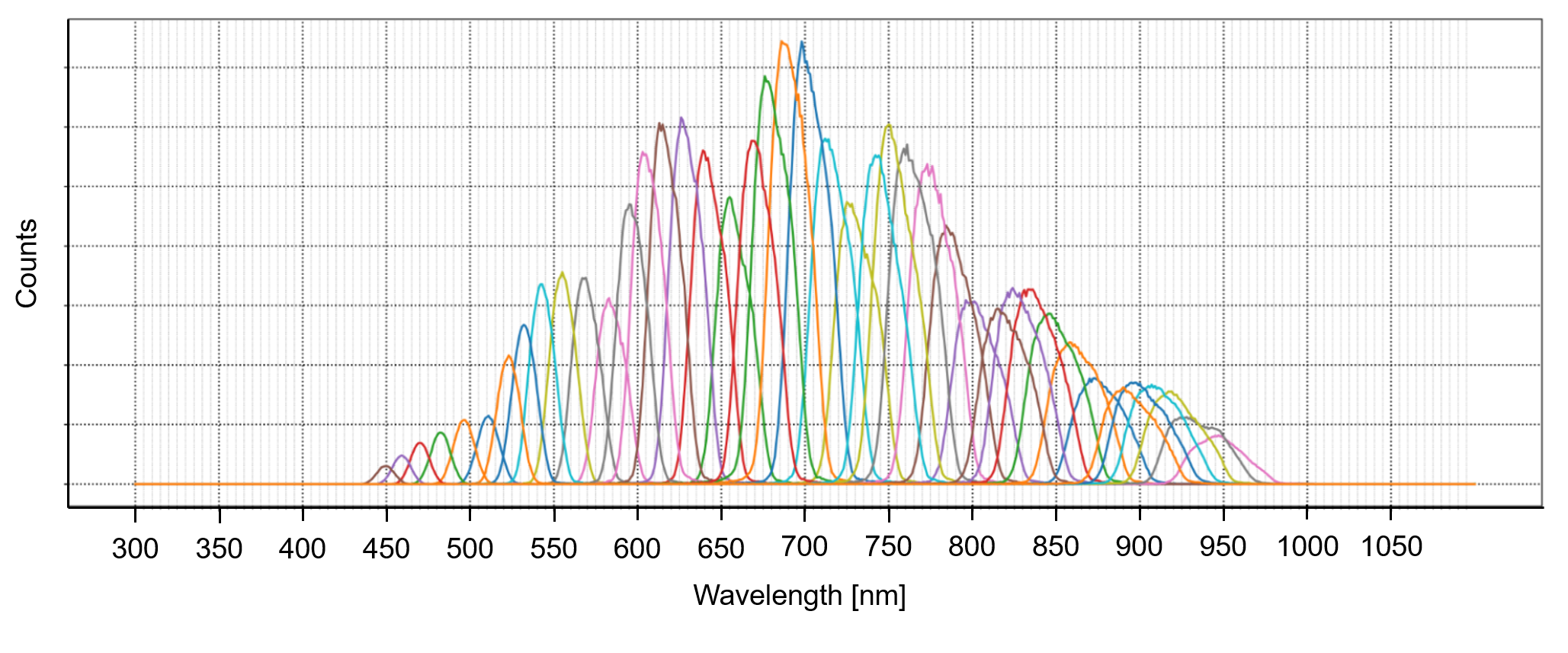}
\caption{
Measurement of the counts using a halogen illumination with the sensor and the filter used. The brightness used for the respective channel is not identical, but it is clearly recognisable that the FWHM increases for longer wavelengths due to dispersion.    
}
\label{fig:channel_overview}
\end{figure*}
The spectral resolution and  measurable wavelength range of the hyperspectral dermatoscope are crucial factors determined by its intended application. Spectral resolution refers to the ability of the dermatoscope to distinguish between different wavelengths. Higher spectral resolution facilitates finer discrimination between wavelengths, thus aiding in  identifying subtle variances in skin lesions. Better spectrally resolved reflectance measurement enables a better estimation of the tissue composition, like skin chromophores and biological parameters of the skin \cite{Halicek2019-zl}. \\ \newline
Achieving a broad range of applications necessitates a measurable range covering the visible spectrum and extending into the near-infrared spectrum (as far as possible). The inclusion of near-infrared wavelengths allows for the detection of reflections from deeper tissue layers \cite{Bashkatov2005-mg}, enhancing diagnostic capabilities. Key spectral landmarks, such as the absorption peak of melanin at 420 nm, the hemoglobin absorption peaks at 540 nm and 575 nm, and the  absorption peak of water at 970 nm \cite{Arimoto2005-fo, Bashkatov2005-mg}, underscore the importance of spanning a measurable spectrum from approximately 400 nm to 1000 nm. \\ \newline
A spectral resolution of around 10 nm would be desirable for resolving fine spectral features of skin lesions. However, achieving this resolution involves a trade-off between spectral resolution, spectral range, portability and spatial resolution. Opting for a high spectral resolution may result in a smaller wavelength range, poor portability (due to the size of the camera) or  lower spatial resolution. One challenge in extending the wavelength range into the near-infrared is ensuring high sensor sensitivity across this entire spectrum. To overcome this challenge, the optics and sensor of the dermatoscope must be specifically designed to accommodate and effectively capture this extended range.  \\ \newline
On the basis of the constraints stated above, we opted for an Ultris SR5 hyperspectral camera from Cubert GmbH, Ulm, Germany, which was tailored to our specific requirements. The acquisition method of the camera is a snapshot method based on light field technology. The camera uses an optical bandpass filter and a micro lens array to perform spectrally resolved measurements. The filter was customised to cover the desired wavelength range as closely as technically possible. The measurable wavelength range of the camera is 450 to 950 nm. This covers the visible range and parts of the near-infrared range. The camera has 51 channels, resulting in a spectral resolution of around 10 nm. A measurement of the combination of camera and the customised filter for the individual channels is given in Figure \ref{fig:channel_overview}. A halogen lamp serves as the light source for the measurement of the sensor response. The brightness of the lamp is not identical for the different channels. However, the measurement shows that the customised filter enables our desired wavelength range. The sensor installed in the camera is the Sony IMX264. The data cubes have a bit depth of 12 bits. Table \ref{table:imaging_techniques} provides an overview of the devices described in section \ref{sec:background} as well as an overview of the properties of the Hyperscope developed by us. The Hyperscope with its exceptionally high spatial and spectral resolution, especially notable for  hand-held devices utilizing the snapshot method. Furthermore, it covers an extensive wavelength range compared to many other devices on the market. One of its stand out features is its remarkable pixel density, which is especially relevant for imaging skin lesions with precision and detail.
\begin{table}[ht]
\centering
\caption{Summary of various hyperspectral imaging devices used for capturing images of skin lesions. If the spatial resolution or the recorded area is not specified, we have entered n.a. in the corresponding columns. }
\label{table:imaging_techniques}
\begin{tabular}{@{}>{\raggedright\arraybackslash}p{2cm} >{\raggedright\arraybackslash}p{2cm} >{\centering\arraybackslash}p{2cm} >{\centering\arraybackslash}p{2cm} >{\centering\arraybackslash}p{2cm} >{\centering\arraybackslash}p{2.2cm} >{\centering\arraybackslash}p{1.8cm}@{}}
\toprule
\textbf{Authors} & \textbf{Imaging Technique} & \textbf{Device Type} & \textbf{Wavelength Range} & \textbf{Number of Bands} & \textbf{Spatial Resolution} & \textbf{Pixel/mm\(^2\)} \\
\midrule
Leon et al. 2020 & Snapshot & Hand-held & 450-950 nm & 125 & 50x50 & 17.36 \\
\addlinespace
Räsänen et al. 2021 & Scanning & Not specified & 500–850 nm & 76 & 240x320 & 640 \\
\addlinespace
Zherdeva et al. 2016 & Scanning & Fixed scanner & 450-750 nm & 61 & 501x501 & 51.22 \\
\addlinespace
Hosking et al. 2019 & Scanning & Hand-held & 350–950 nm & 21 & n.a. & n.a. \\
\addlinespace
Zherebtsov et al. 2019 & Snapshot & Hand-held & 500-900 nm & 40-60 & 1010x1010 & 159.39 \\
\addlinespace
Lindholm et al. 2022 & Scanning & Hand-held & 475-975 nm & 33 & n.a. & n.a. \\
\addlinespace
Calin et al. 2022 & Scanning & Fixed scanner & 400-800 nm & 205 & n.a. & n.a. \\
\addlinespace
\textbf{Hyperscope} & \textbf{Snapshot} & \textbf{Hand-held} & \textbf{450-950 nm} & \textbf{51} & \textbf{290x275} & \textbf{199.38} \\

\bottomrule
\end{tabular}
\end{table}
\subsection{Light source}
\label{sec:light_source}

\begin{figure*}[!t]
\centering
\includegraphics[width=\textwidth]{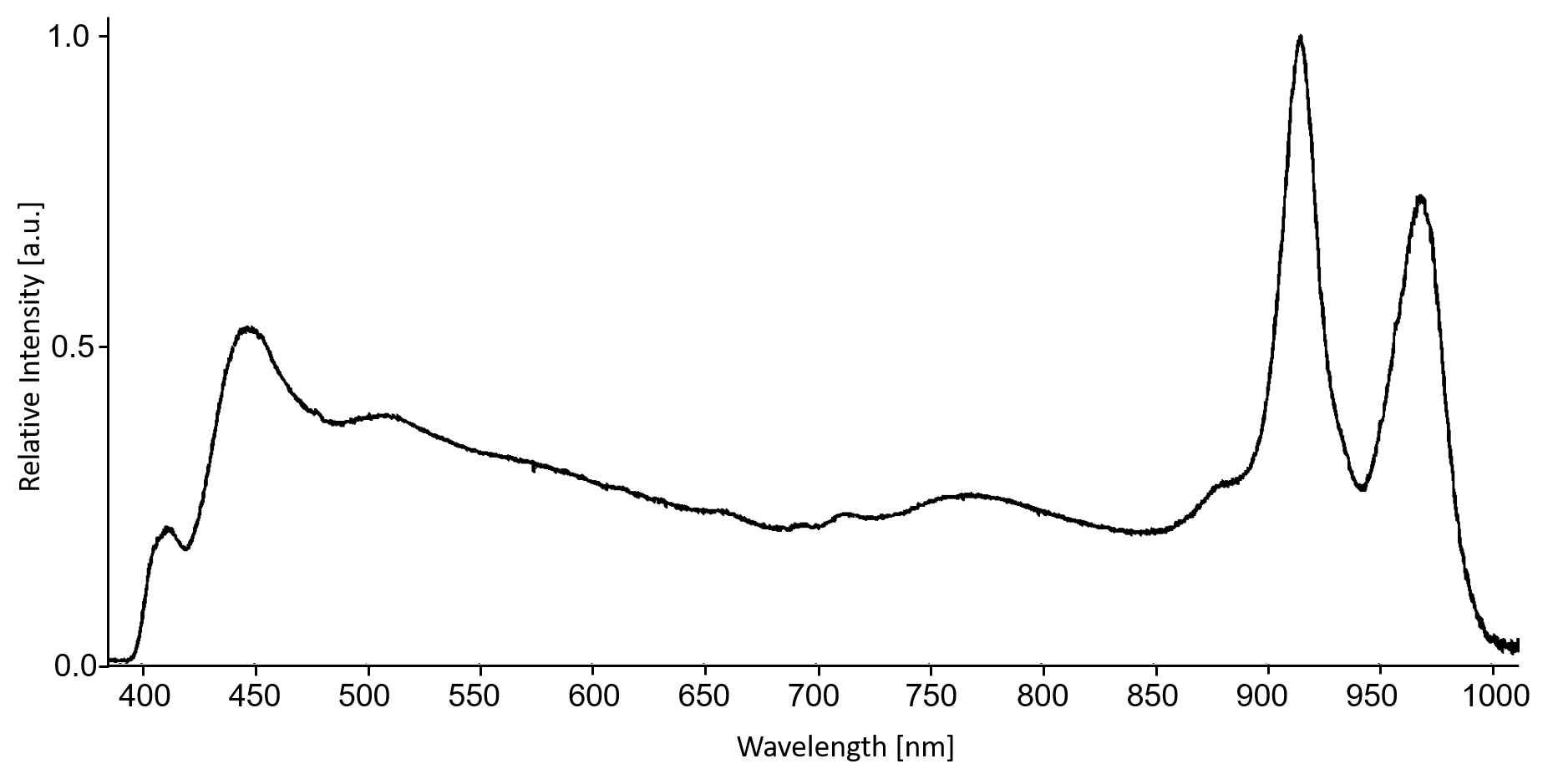}
\caption{
Measured spectrum of the broadband LED lighting used consisting of several white LEDs and two different types of infrared LEDs. The two peaks of the infrared LEDs are at 910 nm and 970 nm.  
}
\label{fig:lighting}
\end{figure*}

To achieve a high-quality signal-to-noise ratio (SNR), it is essential to use a light source with a wide bandwidth. This wide bandwidth ensures uniform illumination across the entire measurable spectrum defined in section \ref{sec:spectral_properties}. Uniform illumination is key to preventing weak signals in certain wavelength ranges, which can lead to noisy images. As the hyperspectral dermatoscope is positioned directly on the patient’s skin, it must be ensured that the heat generated by the illumination cannot cause any harm to the patient. Due to the high efficiency of light emitting diodes (LEDs) compared to incandescent light sources, we choose an LED light source for the Hyperscope developed by us. This minimizes the risk of discomfort or harm to patients due to excessive heat. Furthermore, LEDs can be designed to cover a broad spectrum, encompassing even the near-infrared wavelengths essential for analyzing deep tissue layers. Additionally, LED lighting offers the advantage of compactness, making it favorable for the integration into a hand-held dermatoscope, typically in the form of a ring light or similar design. The illumination profile of the light source depends on the distance to the skin. Depending on the working distance, it may be necessary to use a light guide to ensure homogeneous illumination of the FoV. \\ \newline 
Using  polarised light can offer advantages for imaging by suppressing specular reflections. However, this requires both polarised light and a polarisation filter for the camera, which limits the amount of light hitting the sensor, consequently leading to a longer exposure time. Thus, careful consideration must be given to the selection of the lighting and optics.  \\ \newline
For the Hyperscope, we selected a broadband LED lighting for illumination,  enabling skin illumination across the entire spectrum from 450 nm to 950 nm. For this purpose, several white LEDs were used and combined with two types of infrared LEDs with a peak wavelength of 910 nm and 970 nm. To achieve homogeneous illumination, a light guide was used, ensuring uniform distribution and mixing of the light  the entire FoV. The spectrum of the lighting setup can be seen in Figure \ref{fig:lighting}. In order to compensate for the reduced sensitivity of the CMOS sensor in the infrared range \cite{Yokogawa2017-hi}  and thus maintain a good SNR even for longer wavelengths, high power infrared LEDs were selected. Additionally, the optics and light source are surrounded by a cone which carries the contact plate and efficiently suppresses scattered light from external sources, so that only the light from the designated light source is measured.

\subsection{Optics}
\label{sec:optics}
The optics for the dermatoscope should be selected in such a way that the selected FoV is completely mapped onto the camera sensor. At the same time, the focal length of the lens should match with the working distance, the FoV and the sensor size. Furthermore, the selected lens has to be compatible with the camera mount and should minimise aberrations and provide high transmissivity in the near-infrared range. \\ \newline
The camera we have chosen offers the use of a relay lens, which permits the use of any C-mount compatible lenses. For our Hyperscope, we opted for a 16mm f/1.8, selected for its ability to effectively cover the entire $2$cm$ \times 2$cm inspection area. Attached to the relay lens, it offers exceptional light transmission within the 400-1000 nm range, making it an ideal choice for our application. Additionally, the design allows for easy swapping of lenses. This means that if there is a need for more detailed examination of a smaller area, a lens with a longer focal length can be seamlessly integrated, providing increased magnification for examining smaller skin lesions. The working distance, defined as the distance between the skin surface to be examined and the lens surface, is set at 56 mm.
 \subsection{General design and usability} 
\label{sec:usability}
\begin{figure*}[!t]
\centering
\includegraphics[width=\textwidth]{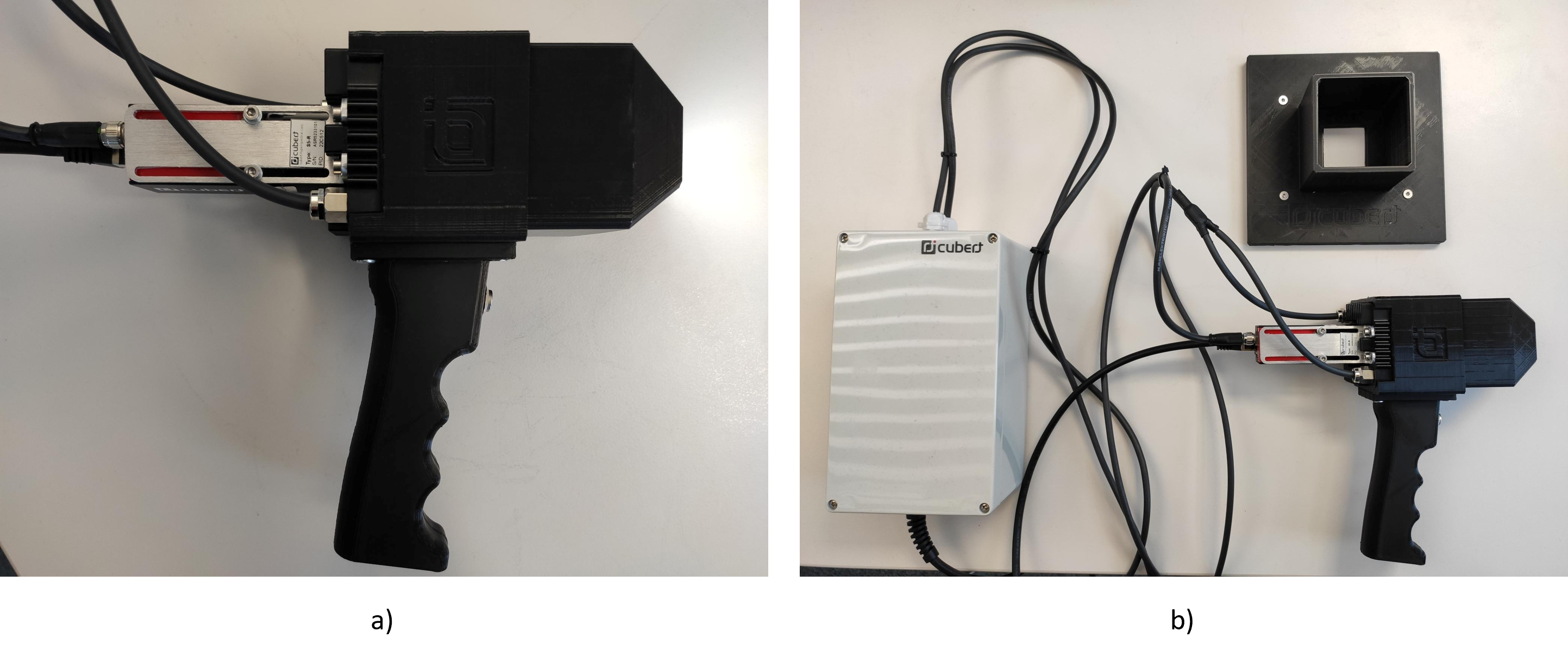}
\caption{
a) Close-up of the developed hyperspectral dermatoscope. b) Image of the entire setup. On the left is the control and power supply for the lighting and camera. The docking station with calibration target can be seen at the top right.  
}
\label{fig:Hyperscope}
\end{figure*}

In designing our hyperspectral dermatoscope, we meticulously considered the requirements essential for its successful integration into clinical environments. Key among these is the necessity for the device to effortlessly blend into the clinical workflow, ensuring it can be utilised as a hand-held tool to provide immediate, on-the-spot diagnostics. One essential feature is a live view capability, enabling practitioners to adjust the position of the dermatoscope in real-time, thereby ensuring the acquisition of optimal images. Additionally, the usability of the software is a crucial consideration, requiring a design that is not only straightforward for clinicians to navigate but also facilitates easy input and management of patient information. \\ \newline
To address these requirements, the Hyperscope is a hand-held device equipped with a customisable button inserted into the grip, facilitating skin measurements by pressing a simple button. A docking station, featuring an integrated calibration target, ensures the device is always primed for accurate imaging upon undocking. The cornerstone of our solution is the custom Python software, which not only offers a live view function for precise device positioning but also simplifies the process of adding and managing patient information, thus enabling a seamless data collection process. This software design prioritises ease of use without compromising on the depth of functionality, ensuring that clinicians can utilise the Hyperscope with minimal learning curve. The Hyperscope, depicted in Figure \ref{fig:Hyperscope}, features a square contact surface using medical certified glass and a 3D-printed housing to eliminate stray light.

\section{Data collection and sample analysis}
\label{sec:data_collection}
To assess the suitability of the Hyperscope for its intended use, we conducted an exemplary data collection as well as preliminary data analysis of 160 collected samples. The focus is on exploring various aspects of the spectral signatures extracted from the HSIs that were collected.
\subsection{Data collection process}
The HSIs were acquired in a controlled laboratory environment by a board certified dermatologist with about 5 years of experience, using our developed prototype. The images were obtained from 15 voluntary participants within our research group. All patients provided informed written consent and the work was performed in accordance with the Declaration of Helsinki. \\ \newline
The data collection process involved a calibration step of the measurements. In order to calculate the reflectance of the skin from the raw images $X$, a white reference $W$ and a dark reference $D$ was obtained. The dark reference image $D$ was obtained by deactivating the light source of the device while it was inserted into the docking station (Figure \ref{fig:Hyperscope}). Conversely, the white reference $W$ was gathered by taking an image of the calibration target inside the docking station. For the recording of the two reference measurements $D$ and $W$, the mean value of 100 recordings was calculated per case. The reflectance $R$ was then derived using equation (\ref{eq:calib}):
\begin{equation}
\label{eq:calib}
R = \frac{X - D}{W - D}
\end{equation}
Images were taken of each patient’s arms, legs, face, neck and hands. Any pigmented lesions were recorded, and if none were present on the respective body part, an image of the skin was still taken. In addition, any large pigmented lesions
on the torso or abdomen were recorded. A total of 160 images were recorded, 91 of which contained a pigmented lesion. For further analyses, a point annotation was performed for each measurement. This involved selecting a representative point in the image for the ’Lesion’ or ’Skin’ class. The center of a lesion served as the annotation target for the ’Lesion’ class, while annotations for the ’Skin’ class were chosen as centrally as possible.

\subsection{Data analysis}
\begin{figure*}[!htb]
\centering
\includegraphics[width=0.9\textwidth]{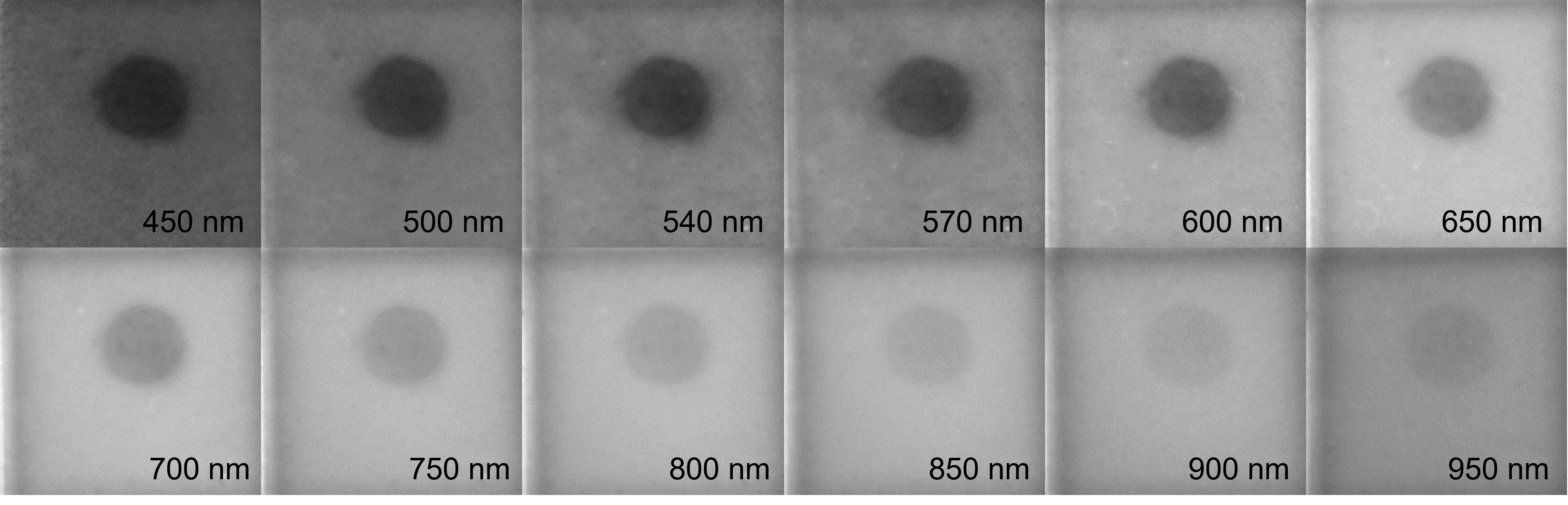}
\caption{
A pigmented lesion located on the torso viewed at different wavelengths. The images show the lesion with steps of 50 nm. Except for 3 and 4, which show the wavelengths 540 nm and 570 nm respectively, which represent the hemoglobin absorption peaks. With increasing wavelength in the visible range, the increased reflectance can be clearly recognised in the image. 
}
\label{fig:lesion_images}
\end{figure*}
\begin{figure*}[!htb]
\centering
\includegraphics[width=\textwidth]{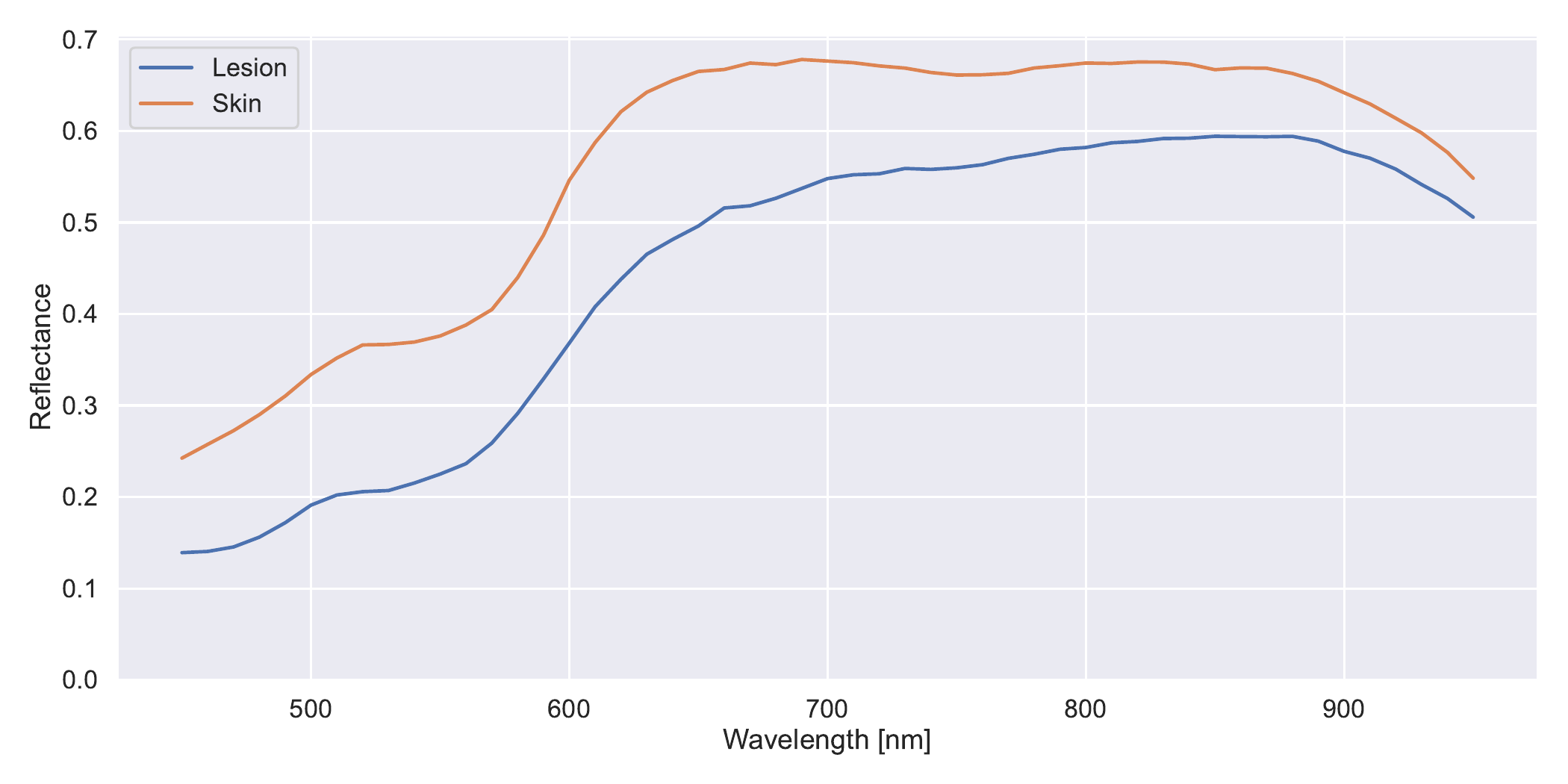}
\caption{
Median reflectance of the 'Skin' and 'Lesion' class of all body parts and all patient ids. For each measurement, the mean value was determined over an area of $3\times3$ pixels around the annotated location. The median is shown here since it preserves spectral features better than the mean. 
}
\label{fig:median_reflectance}
\end{figure*}
\begin{figure*}[!htb]
\centering
\includegraphics[width=\textwidth]{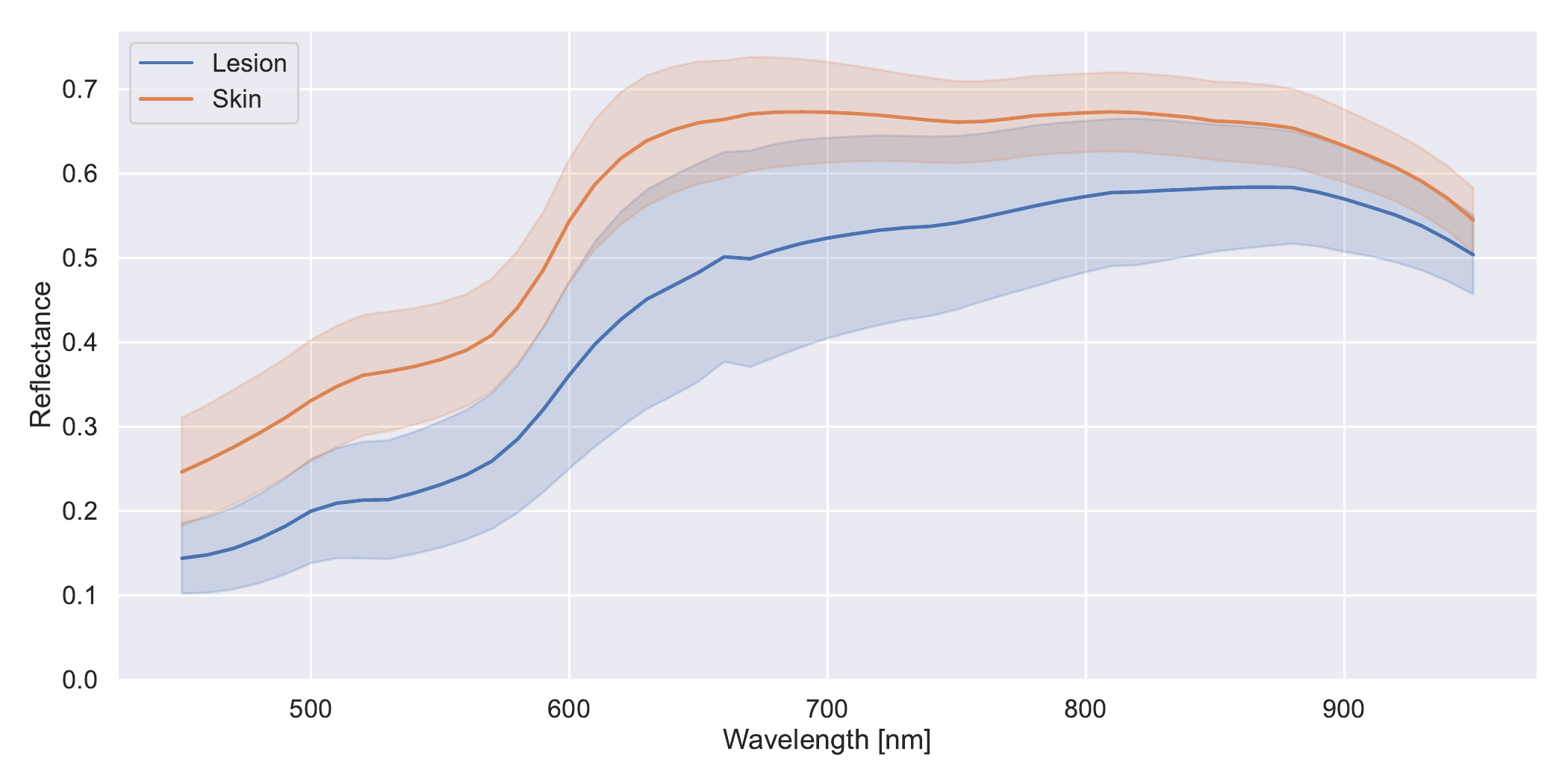}
\caption{
Mean reflectance and standard deviation of the 'Skin' and 'Lesion' class for all patients and bodyparts. For each measurement, the mean value was determined over an area of $3\times3$ pixels around the annotated location. The reflectance of pigmented lesions in the visible area is significantly lower than that of normal skin.
}
\label{fig:skin_vs_lesion}
\end{figure*}
\begin{figure*}[!ht]
\centering
\includegraphics[width=\textwidth]{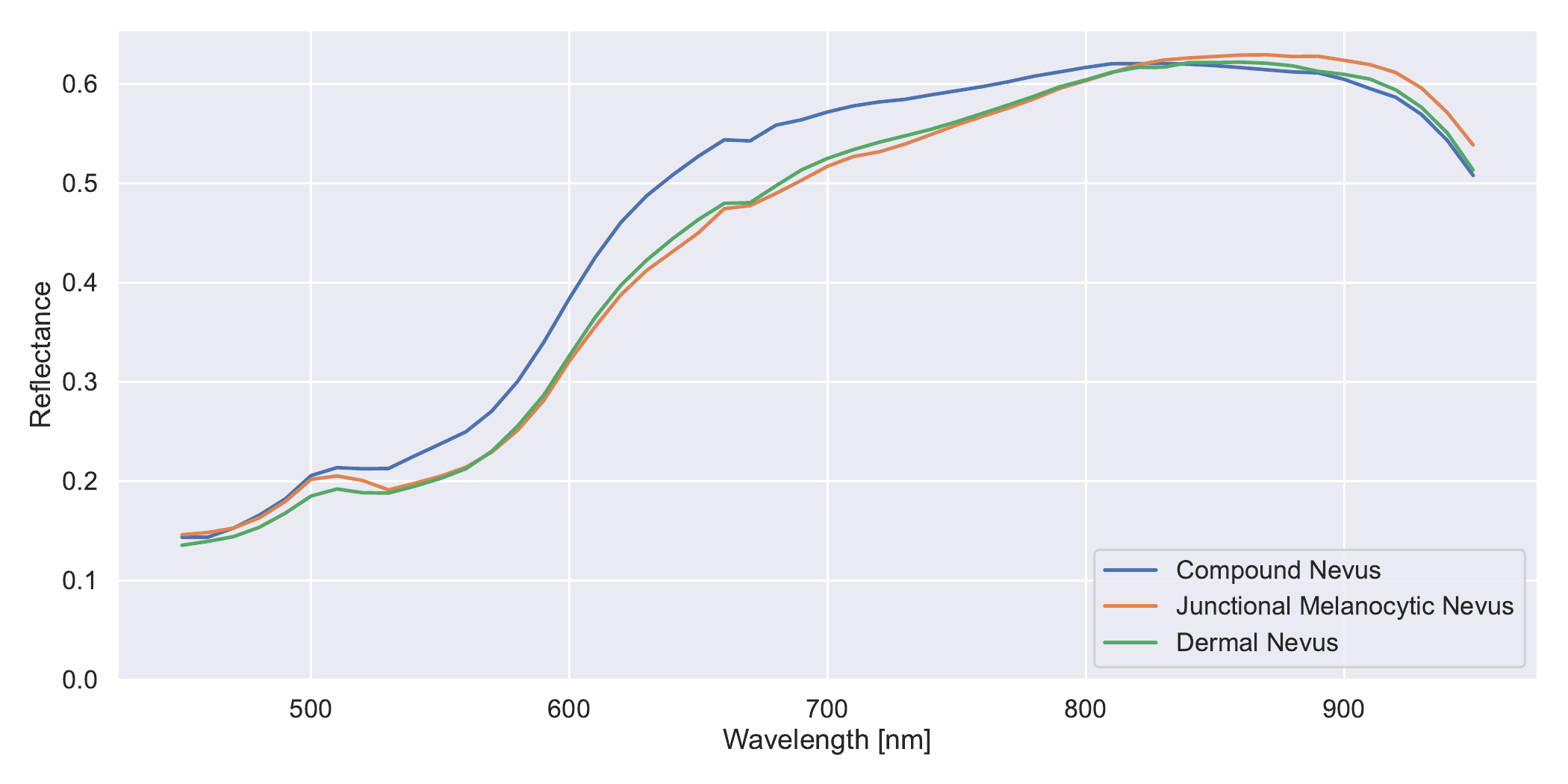}
\caption{
Mean reflectance spectra of multiple pigmented lesions of a single patient stratified by histologic type. 
}
\label{fig:histo_pattern}
\end{figure*}

The collected measurements are processed and analysed using Python and the Software development kit (SDK\footnote{https://github.com/cubert-hyperspectral/cuvis.sdk}) of the camera. To ensure a more accurate representation of the data, the mean value of  $3 \times 3$ pixel area around the point annotation is used in all analyses. Using  the neighbouring pixels allows for a precise annotation while simultaneously  some variability of the data. Figure \ref{fig:median_reflectance} shows the median reflectance for the ’Skin’ and ’Lesion’ classes. The median was determined using all body parts and patients. Compared to the mean value, which can be seen in Figure \ref{fig:skin_vs_lesion},  the median better preserves spectral features. \\ \newline
A visual comparison with the spectra measured by a spectrometer (e.g., \cite{Cooksey2015-ak}) in the range from 450 nm to 950 nm reveals that the data measured by our hyperspectral dermatoscope (the Hyperscope) reproduce a comparable reflectance behaviour. Our reflectance spectrum shows less pronounced minima and maxima, which can be explained by a lower spectral resolution. The determined reflectance clearly shows the absorption peak of hemoglobin at 540 nm and the sharp drop in reflectance attributed to the water absorption peak at 970 nm \cite{Arimoto2005-fo, Bashkatov2005-mg}, which, however, lies beyond our recorded wavelength range. \\ \newline
A sample taken with our hyperspectral dermatoscope (the Hyperscope) can be seen in Figure \ref{fig:lesion_images}. This figure illustrates the reflectance of a pigmented skin lesion located on the torso at different wavelengths. It is evident that the lesion reflects significantly more radiation in the NIR range compared to the visible range. This characteristic is also observable in Figure \ref{fig:skin_vs_lesion}, where the reflectance of the lesion and normal skin converge in the NIR range. This behaviour is strongly attributed to the decreasing absorption of melanin, which is typically found in higher concentrations in pigmented lesions, in the NIR range. \\ \newline
Despite a homogeneous cohort in terms of age and skin phototype, the standard deviation for both the ’Lesion’ class and the ’Skin’ class shows substantial variability in skin reflectance. This is likely due to the optical properties of the skin such as scattering and absorption characteristics. These characteristics vary both between individuals and between the body parts of a single individual. 
\begin{figure*}[!htb]
\centering
\includegraphics[width=\textwidth]{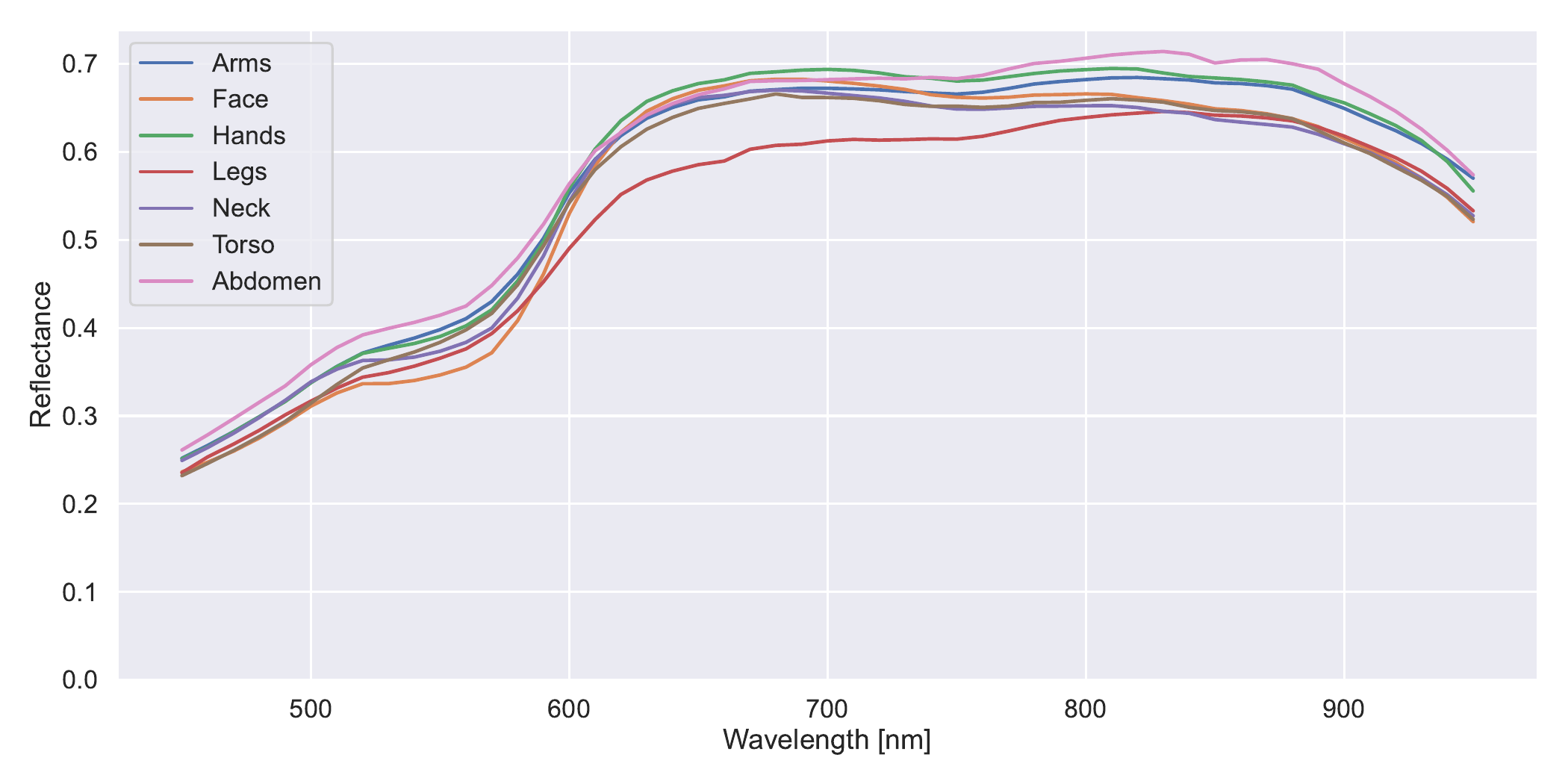}
\caption{
Mean reflectance of the 'Skin' class stratified by body part for all patients. For each measurement, the mean value was determined over an area of $3\times3$ pixels around the annotated location. It can be observed that different parts of the body have different reflectance due to differing physiological parameters. 
}
\label{fig:skin_bodyparts}
\end{figure*}
Figure \ref{fig:skin_bodyparts} shows the mean reflectance of the ’Skin’ class, averaged over all persons, of different body parts. These differences are due to different biological parameters such as collagen structure, melanin concentration or thickness of the different skin layers.\\ \newline
The notably lower reflectance of the skin in the ’Legs’ class can possibly be attributed to increased body hair. This explanation is supported by the behaviour observed in the NIR range, as hair tends to be almost transparent to infrared radiation. This characteristic is shown in the graph by the convergence of the reflectance of the ’Legs’ class with the reflectance of other body parts. 
\begin{figure*}[!htb]
\centering
\includegraphics[width=\textwidth]{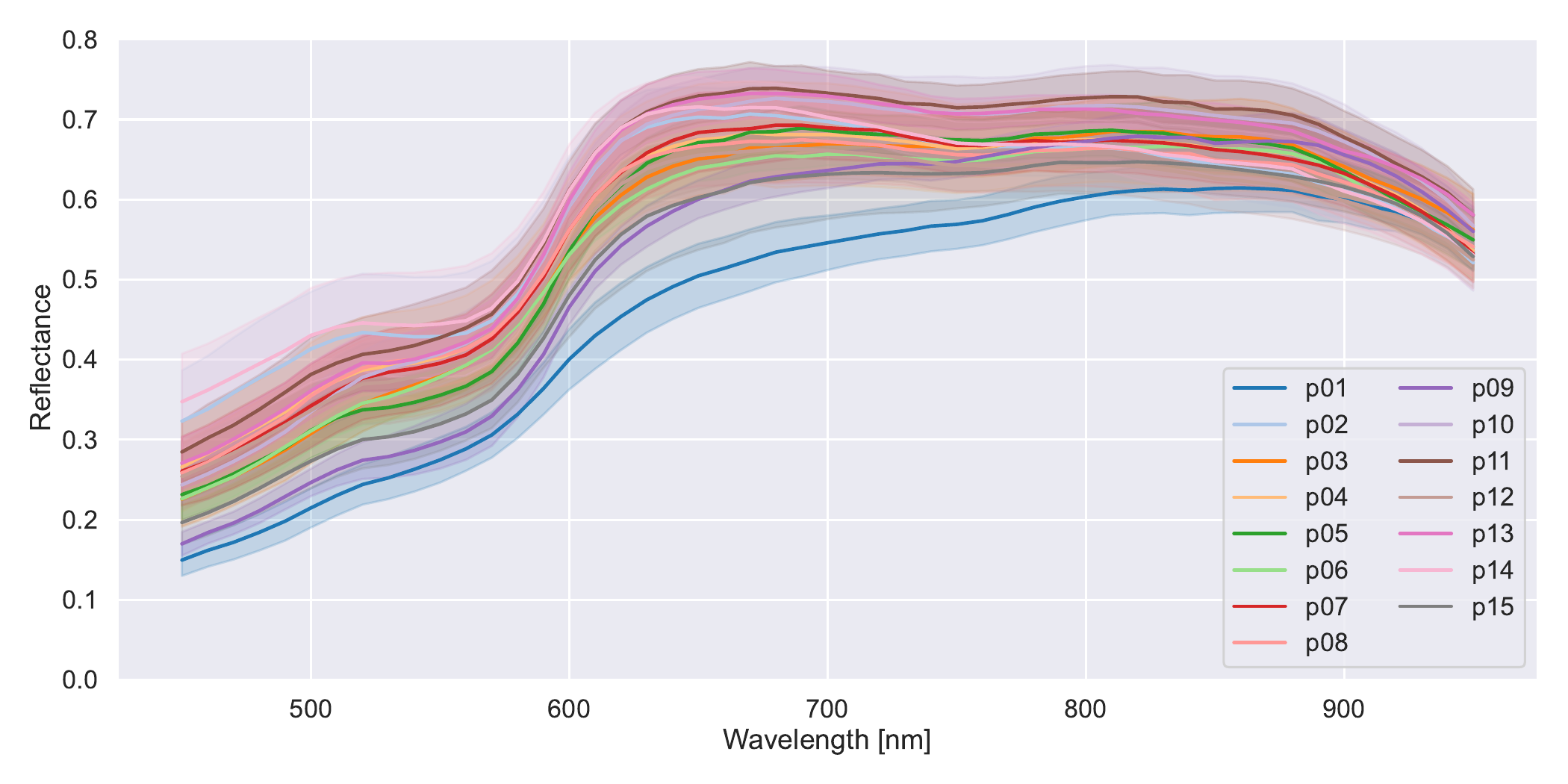}
\caption{
Mean reflectance and standard deviation of the 'Skin' class stratified by patient id averaged over all body parts. For each measurement, the mean value was determined over an area of $3\times3$ pixels around the annotated location. It can be observed that the skin of the 15 patients has a unique spectral signature across the different parts of the body.  
}
\label{fig:patient_signature}
\end{figure*}
The high variability of the reflectance of the skin is also shown in Figure \ref{fig:patient_signature}. Here, the mean reflectance for the 'Skin' class is averaged over all body parts stratified by patient id.  Moreover, distinct differences in reflectance behaviour, especially in the visible area, are observed between individuals. These differences are mainly attributed to differences in phototype. \\ \newline
Furthermore, we analysed the spectral signatures of nevi with different dermoscopic patterns and histopathological types. Melanocytic lesions present dermoscopic patterns and features that correlate to histological morphology \cite{Kasuya2019-dy}. Against this background, we recorded dermoscopic images of different types of melanocytic nevi located on the trunk of a single volunteer (male, 27 years old) and labelled  their histopathological type and corresponding pattern together with a board-certified dermatologist, in order to investigate whether there is a correlation between dermoscopic patterns and spectral signature. The spectral signatures are visualised in Figure \ref{fig:derma_patterns}. Example dermoscopic images of the lesions recorded are shown in the appendix in Figure \ref{fig:derma_images}.
\begin{figure*}[!ht]
\centering
\includegraphics[width=\textwidth]{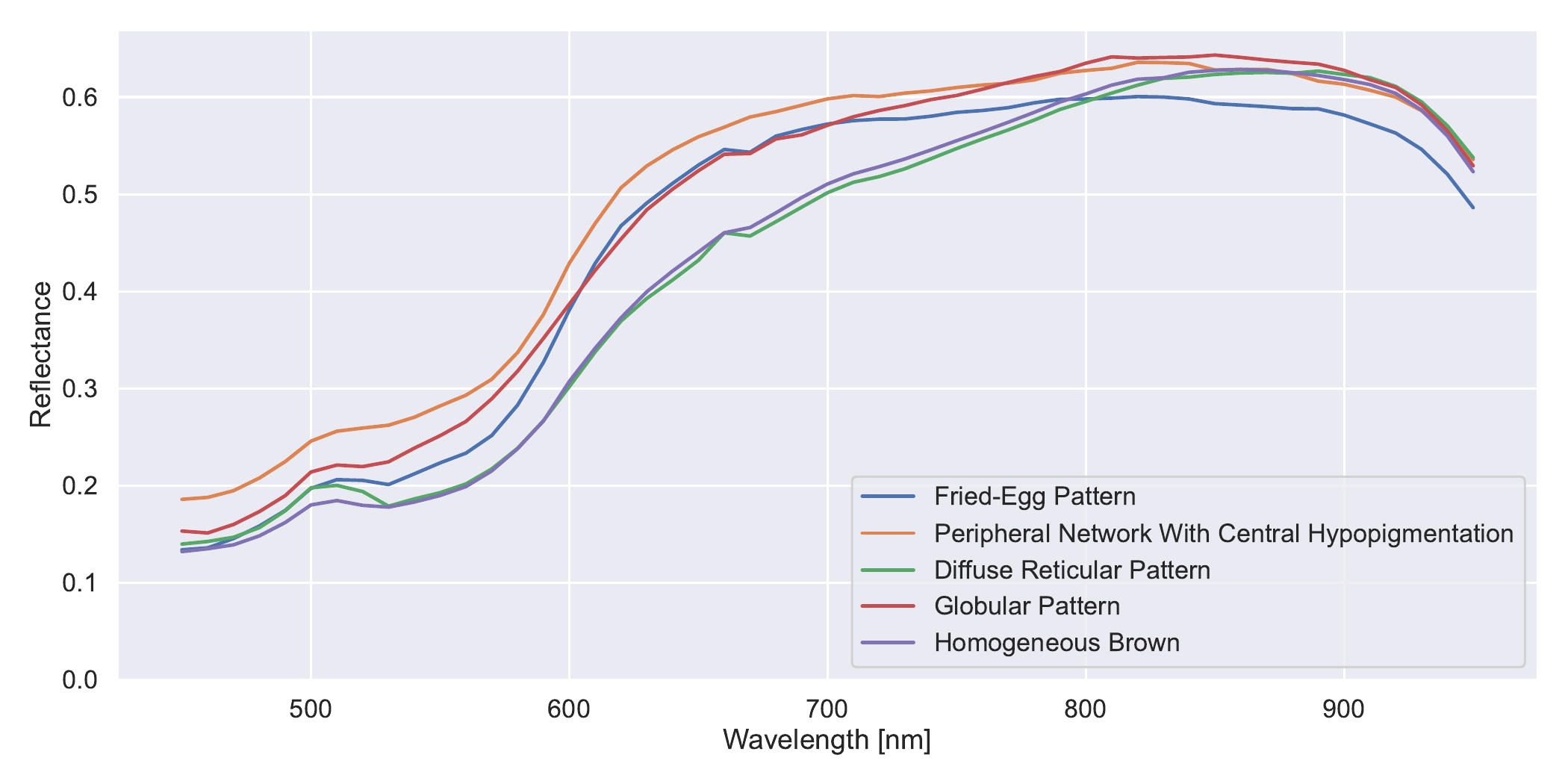}
\caption{
Mean reflectance spectra of multiple pigmented lesions of a single patient stratified by dermoscopic pattern. 
}
\label{fig:derma_patterns}
\end{figure*}
Benign melanocytic patterns present in our sample were reticular (diffuse reticular, peripheral network with central hypopigmentation), globular (cobblestone, fried-egg pattern) and homogeneous brown \cite{Gulia2012-ta}. Recognising relevant pigment variance, we opted to separate lesions into different groups based on dermoscopic patterns, particularly focusing on peripheral network with central hypopigmentation and fried-egg pattern. Subsequently, we plotted reflectance curves stratified by dermoscopic patterns to explore how skin chromophores, especially melanin, can interfere in the reflectance curve. Our  observations revealed a distinct separation among three reflectance curves from 450 nm until 500 nm: peripheral network with central hypopigmentation, globular pattern and homogeneous brown pattern. However, the diffuse reticular and fried-egg patterns exhibit overlap within this range. \\ \newline
The reflectance curve of the diffuse reticular pattern fluctuates between the homogeneous brown and fried-egg pattern. Below 500 nm, it overlaps with fried-egg, while above 530 nm, it overlaps with the homogeneous brown pattern. The behaviour below 500 nm may be attributed to a reticular component in the fried-egg pattern with an important component of melanin absorbance. Conversely, the fried-egg pattern could be influenced by haemoglobin absorbance above 530 nm. Central hypopigmentation notably influences the reflectance, maintaining an orange curve with the highest reflectance up to 750 nm. The fried-egg pattern presents the lowest reflectance in NIR, possibly due to a heightened influence of haemoglobin and water absorption, given the skin projection of the globular component and lower melanin concentration. \\ \newline
We aimed to investigate whether different histological morphologies of melanocytic nevi could influence lesion reflectance. In terms of skin layers involved, melanocytic nevi are histologically classified in junctional (epidermis), compound (epidermis and dermis) and dermal nevus (dermis). The measured reflectance, stratified by histological pattern, is presented in Figure  \ref{fig:histo_pattern}. \\ \newline
We observed that junctional nevi fluctuate with relative high reflectance, overlapping with compound nevi below 500 nm. However, over 530 nm, they assume a relative low reflectance overlapping with dermal nevi. Between 540 and 820 nm, dermal and junctional nevi behave similarly, suggesting that in this range hemoglobin absorption is more relevant than melanin. Compound nevi present the highest reflectance between 500 and 800 nm with a drop in NIR range. This observation could also be justified by hemoglobin absorbance rather than melanin. Skin layer involvement does not appear to be the primary influence on the reflectance curve.
\section{Conclusion}
In conclusion, our study validates the capability of the developed hyperspectral dermatoscope (the Hyperscope) to accurately replicate skin reflectance spectra as documented in existing literature, demonstrating its utility in a clinical setting. The design of the device, leveraging the snapshot method, ensures both high spatial and spectral resolutions, enabling the application of advanced deep learning models for skin analysis. Preliminary findings suggest the potential of our device to discern between physiological characteristics of skin lesions, offering a pathway towards the biologically-informed classification of such conditions. The Hyperscope is setting a foundation for future research in non-invasive skin lesion diagnosis and potentially transforming dermatological practice. \\ \newline
As we continue to refine our technology, we anticipate broader applications, including phototype classification, malignancy detection in pigmented lesions and the estimation of skin age, all through spectral signature analysis. This endeavor not only marks a notable step towards enhancing dermatological diagnostic processes but also opens up new avenues for personalized dermatology care. In future projects, the Hyperscope will be distributed in several clinics to collect data within an real-life clinical environment. This data can be used to develop algorithms that are capable of distinguishing malignant from benign lesions using spectral and spatial characteristics. We remain confident that a richer data base
compared to conventional dermoscopic images will substantially improve specificity and sensitivity in the field of skin cancer diagnostics, boosting the urgently needed skin cancer recognition at an early stage.

\section*{Acknowledgements}
This study was funded by the Federal Ministry of Health, Berlin, Germany (grant: Skin Classification Project 2) and the Ministry of Health, Social Affairs and Integration Baden-Württemberg, Stuttgart, Germany (grant: sKIn). Grant holder in both cases: Titus J. Brinker, German Cancer Research Center, Heidelberg, Germany.
C.N.G. was supported by the DKFZ Clinician Scientist Program and the Dieter Morszeck Foundation.

%Bibliography
\bibliographystyle{unsrt}  
\bibliography{references}  

\begin{thebibliography}{10}

\bibitem{SkinCancer}
Wcrf international - skin cancer statistics.
\newblock \url{https://www.wcrf.org/cancer-trends/skin-cancer-statistics/}.
\newblock Accessed: 2023-12-13.

\bibitem{Urban2021-ty}
Katelyn Urban, Sino Mehrmal, Prabhdeep Uppal, Rachel~L Giesey, and Gregory~R Delost.
\newblock The global burden of skin cancer: A longitudinal analysis from the global burden of disease study, 1990--2017.
\newblock {\em JAAD International}, 2:98--108, March 2021.

\bibitem{Tripp2016-am}
Mary~K Tripp, Meg Watson, Sophie~J Balk, Susan~M Swetter, and Jeffrey~E Gershenwald.
\newblock State of the science on prevention and screening to reduce melanoma incidence and mortality: The time is now.
\newblock {\em CA Cancer J. Clin.}, 66(6):460--480, November 2016.

\bibitem{American_Academy_of_Dermatology_Ad_Hoc_Task_Force_for_the_ABCDEs_of_Melanoma2015-tw}
{American Academy of Dermatology Ad Hoc Task Force for the ABCDEs of Melanoma}, Hensin Tsao, Jeannette~M Olazagasti, Kelly~M Cordoro, Jerry~D Brewer, Susan~C Taylor, Jeremy~S Bordeaux, Mary-Margaret Chren, Arthur~J Sober, Connie Tegeler, Reva Bhushan, and Wendy~Smith Begolka.
\newblock Early detection of melanoma: reviewing the {ABCDEs}.
\newblock {\em J. Am. Acad. Dermatol.}, 72(4):717--723, April 2015.

\bibitem{Kittler2002-vb}
H~Kittler, H~Pehamberger, K~Wolff, and M~Binder.
\newblock Diagnostic accuracy of dermoscopy.
\newblock {\em Lancet Oncol.}, 3(3):159--165, March 2002.

\bibitem{Vestergaard2008-yl}
M~E Vestergaard, P~Macaskill, P~E Holt, and S~W Menzies.
\newblock Dermoscopy compared with naked eye examination for the diagnosis of primary melanoma: a meta‐analysis of studies performed in a clinical setting.
\newblock {\em Br. J. Dermatol.}, 159(3):669--676, September 2008.

\bibitem{SkinCancer2}
National cancer institute - cancer stat facts: Melanoma of the skin.
\newblock \url{https://seer.cancer.gov/statfacts/html/melan.html}.
\newblock Accessed: 2023-01-05.

\bibitem{Mayer1997-ke}
J~Mayer.
\newblock Systematic review of the diagnostic accuracy of dermatoscopy in detecting malignant melanoma.
\newblock {\em Med. J. Aust.}, 167(4):206--210, August 1997.

\bibitem{Augustin1999-bp}
M~Augustin, I~Zschocke, N~Godau, A~Buske-Kirschbaum, M~Peschen, B~Sommer, and G~Sattler.
\newblock Skin surgery under local anesthesia leads to stress-induced alterations of psychological, physical, and immune functions.
\newblock {\em Dermatol. Surg.}, 25(11):868--871, November 1999.

\bibitem{Vaidya2019-bk}
Toral~S Vaidya, Shoko Mori, Stephen~W Dusza, Anthony~M Rossi, Kishwer~S Nehal, and Erica~H Lee.
\newblock Appearance-related psychosocial distress following facial skin cancer surgery using the {FACE-Q} skin cancer.
\newblock {\em Arch. Dermatol. Res.}, 311(9):691--696, November 2019.

\bibitem{Heibel2020-ng}
Haley~D Heibel, Leah Hooey, and Clay~J Cockerell.
\newblock A review of noninvasive techniques for skin cancer detection in dermatology.
\newblock {\em Am. J. Clin. Dermatol.}, 21(4):513--524, August 2020.

\bibitem{Soglia2022-vf}
Simone Soglia, Javiera P{\'e}rez-Anker, Nelson Lobos~Guede, Priscila Giavedoni, Susana Puig, and Josep Malvehy.
\newblock Diagnostics using {Non-Invasive} technologies in dermatological oncology.
\newblock {\em Cancers}, 14(23), November 2022.

\bibitem{Lu2014-wp}
Guolan Lu and Baowei Fei.
\newblock Medical hyperspectral imaging: a review.
\newblock {\em J. Biomed. Opt.}, 19(1):10901, January 2014.

\bibitem{Halicek2019-zl}
Martin Halicek, Himar Fabelo, Samuel Ortega, Gustavo~M Callico, and Baowei Fei.
\newblock {In-Vivo} and {Ex-Vivo} tissue analysis through hyperspectral imaging techniques: Revealing the invisible features of cancer.
\newblock {\em Cancers}, 11(6), May 2019.

\bibitem{Yoon2022-ea}
Jonghee Yoon.
\newblock Hyperspectral imaging for clinical applications.
\newblock {\em Biochip J.}, 16(1):1--12, March 2022.

\bibitem{Aggarwal2022-ji}
St~Lt~Pushkar Aggarwal and Francis~A Papay.
\newblock Applications of multispectral and hyperspectral imaging in dermatology.
\newblock {\em Exp. Dermatol.}, 31(8):1128--1135, August 2022.

\bibitem{Lindholm2022-az}
Vivian Lindholm, Anna-Maria Raita-Hakola, Leevi Annala, Mari Salmivuori, Leila Jeskanen, Heikki Saari, Sari Koskenmies, Sari Pitk{\"a}nen, Ilkka P{\"o}l{\"o}nen, Kirsi Isoherranen, and Annamari Ranki.
\newblock Differentiating malignant from benign pigmented or {Non-Pigmented} skin {Tumours---A} pilot study on {3D} hyperspectral imaging of complex skin surfaces and convolutional neural networks.
\newblock {\em J. Clin. Med. Res.}, 11(7):1914, March 2022.

\bibitem{Calin2022-im}
Mihaela~Antonina Calin and Sorin~Viorel Parasca.
\newblock Automatic detection of basal cell carcinoma by hyperspectral imaging.
\newblock {\em J. Biophotonics}, 15(1):e202100231, January 2022.

\bibitem{Torti2020-oa}
Emanuele Torti, Raquel Leon, Marco La~Salvia, Giordana Florimbi, Beatriz Martinez-Vega, Himar Fabelo, Samuel Ortega, Gustavo~M Callic{\'o}, and Francesco Leporati.
\newblock Parallel classification pipelines for skin cancer detection exploiting hyperspectral imaging on hybrid systems.
\newblock {\em Electronics}, 9(9):1503, September 2020.

\bibitem{Courtenay2021-uh}
Lloyd~A Courtenay, Diego Gonz{\'a}lez-Aguilera, Susana Lag{\"u}ela, Susana Del~Pozo, Camilo Ruiz-Mendez, In{\'e}s Barbero-Garc{\'\i}a, Concepci{\'o}n Rom{\'a}n-Curto, Javier Ca{\~n}ueto, Carlos Santos-Dur{\'a}n, Mar{\'\i}a~Esther Carde{\~n}oso-{\'A}lvarez, M{\'o}nica Roncero-Riesco, David Hernandez-Lopez, Diego Guerrero-Sevilla, and Pablo Rodr{\'\i}guez-Gonzalvez.
\newblock Hyperspectral imaging and robust statistics in non-melanoma skin cancer analysis.
\newblock {\em Biomed. Opt. Express}, 12(8):5107--5127, August 2021.

\bibitem{Hosking2019-ua}
Anna-Marie Hosking, Brandon~J Coakley, Dorothy Chang, Faezeh Talebi-Liasi, Samantha Lish, Sung~Won Lee, Amanda~M Zong, Ian Moore, James Browning, Steven~L Jacques, James~G Krueger, Kristen~M Kelly, Kenneth~G Linden, and Daniel~S Gareau.
\newblock Hyperspectral imaging in automated digital dermoscopy screening for melanoma.
\newblock {\em Lasers Surg. Med.}, 51(3):214--222, March 2019.

\bibitem{Huang2024-pj}
Hung-Yi Huang, Hong-Thai Nguyen, Teng-Li Lin, Penchun Saenprasarn, Ping-Hung Liu, and Hsiang-Chen Wang.
\newblock Identification of skin lesions by snapshot hyperspectral imaging.
\newblock {\em Cancers}, 16(1), January 2024.

\bibitem{Huang2023-uq}
Hung-Yi Huang, Yu-Ping Hsiao, Arvind Mukundan, Yu-Ming Tsao, Wen-Yen Chang, and Hsiang-Chen Wang.
\newblock Classification of skin cancer using novel hyperspectral imaging engineering via {YOLOv5}.
\newblock {\em J. Clin. Med. Res.}, 12(3), February 2023.

\bibitem{Rasanen2021-px}
Janne R{\"a}s{\"a}nen, Mari Salmivuori, Ilkka P{\"o}l{\"o}nen, Mari Gr{\"o}nroos, and Noora Neittaanm{\"a}ki.
\newblock Hyperspectral imaging reveals spectral differences and can distinguish malignant melanoma from pigmented basal cell carcinomas: A pilot study.
\newblock {\em Acta Derm. Venereol.}, 101(2):adv00405, February 2021.

\bibitem{Leon2020-tr}
Raquel Leon, Beatriz Martinez-Vega, Himar Fabelo, Samuel Ortega, Veronica Melian, Irene Casta{\~n}o, Gregorio Carretero, Pablo Almeida, Aday Garcia, Eduardo Quevedo, Javier~A Hernandez, Bernardino Clavo, and Gustavo M~Callico.
\newblock {Non-Invasive} skin cancer diagnosis using hyperspectral imaging for {In-Situ} clinical support.
\newblock {\em J. Clin. Med. Res.}, 9(6), June 2020.

\bibitem{Gevaux2021-zm}
Lou Gevaux, Jordan Gierschendorf, Juliette Rengot, Marie Cherel, Pierre S{\'e}roul, Alex Nkengne, Julie Robic, Alain Tr{\'e}meau, and Mathieu H{\'e}bert.
\newblock Real-time skin chromophore estimation from hyperspectral images using a neural network.
\newblock {\em Skin Res. Technol.}, 27(2):163--177, March 2021.

\bibitem{Zherebtsov2019-ds}
Evgeny Zherebtsov, Viktor Dremin, Alexey Popov, Alexander Doronin, Daria Kurakina, Mikhail Kirillin, Igor Meglinski, and Alexander Bykov.
\newblock Hyperspectral imaging of human skin aided by artificial neural networks.
\newblock {\em Biomed. Opt. Express}, 10(7):3545--3559, July 2019.

\bibitem{Fabelo2019-ug}
Himar Fabelo, Ver{\'o}nica Meli{\'a}n, Beatriz Mart{\'\i}nez, Patricia Beltr{\'a}n, Samuel Ortega, Margarita Marrero, Gustavo~M Callic{\'o}, Roberto Sarmiento, Irene Casta{\~n}o, Gregorio Carretero, Pablo Almeida, Aday Garc{\'\i}a, Javier~A Hern{\'a}ndez, and Fred Godtliebsen.
\newblock Dermatologic hyperspectral imaging system for skin cancer diagnosis assistance.
\newblock In {\em 2019 {XXXIV} Conference on Design of Circuits and Integrated Systems ({DCIS})}, pages 1--6. IEEE, November 2019.

\bibitem{Vasefi2014-kp}
Fartash Vasefi, Nicholas MacKinnon, Rolf~B Saager, Anthony~J Durkin, Robert Chave, Erik~H Lindsley, and Daniel~L Farkas.
\newblock Polarization-sensitive hyperspectral imaging in vivo: a multimode dermoscope for skin analysis.
\newblock {\em Sci. Rep.}, 4:4924, May 2014.

\bibitem{Zherdeva2016-eb}
Larisa~A Zherdeva, Ivan~A Bratchenko, Oleg~O Myakinin, Alexander~A Moryatov, Sergey~V Kozlov, and Valery~P Zakharov.
\newblock In vivo hyperspectral imaging and differentiation of skin cancer.
\newblock In {\em Optics in Health Care and Biomedical Optics {VII}}, volume 10024, pages 658--665. SPIE, October 2016.

\bibitem{Aloupogianni2021-pk}
Eleni Aloupogianni, Masahiro Ishikawa, Takaya Ichimura, Atsushi Sasaki, Naoki Kobayashi, and Takashi Obi.
\newblock Design of a {Hyper-Spectral} imaging system for gross pathology of pigmented skin lesions.
\newblock {\em Conf. Proc. IEEE Eng. Med. Biol. Soc.}, 2021:3605--3608, November 2021.

\bibitem{Raita-Hakola2022-he}
Anna-Maria Raita-Hakola, Leevi Annala, Vivian Lindholm, Roberts Trops, Antti N{\"a}sil{\"a}, Heikki Saari, Annamari Ranki, and Ilkka P{\"o}l{\"o}nen.
\newblock {FPI} based hyperspectral imager for the complex {Surfaces---Calibration}, illumination and applications.
\newblock {\em Sensors}, 22(9):3420, April 2022.

\bibitem{Aloupogianni2022-ah}
Eleni Aloupogianni, Takaya Ichimura, Mei Hamada, Masahiro Ishikawa, Takuo Murakami, Atsushi Sasaki, Koichiro Nakamura, Naoki Kobayashi, and Takashi Obi.
\newblock Hyperspectral imaging for tumor segmentation on pigmented skin lesions.
\newblock {\em J. Biomed. Opt.}, 27(10), October 2022.

\bibitem{Johnson2007-ym}
William~R Johnson, Daniel~W Wilson, Wolfgang Fink, Mark Humayun, and Greg Bearman.
\newblock Snapshot hyperspectral imaging in ophthalmology.
\newblock {\em J. Biomed. Opt.}, 12(1):014036, 2007.

\bibitem{Hagen2013-zy}
Nathan~A Hagen and Michael~W Kudenov.
\newblock Review of snapshot spectral imaging technologies.
\newblock {\em Organ. Ethic.}, 52(9):090901, September 2013.

\bibitem{Dosovitskiy2020-gr}
Alexey Dosovitskiy, Lucas Beyer, Alexander Kolesnikov, Dirk Weissenborn, Xiaohua Zhai, Thomas Unterthiner, Mostafa Dehghani, Matthias Minderer, Georg Heigold, Sylvain Gelly, Jakob Uszkoreit, and Neil Houlsby.
\newblock An image is worth 16x16 words: Transformers for image recognition at scale.
\newblock October 2020.

\bibitem{He2015-ba}
Kaiming He, X~Zhang, Shaoqing Ren, and Jian Sun.
\newblock Deep residual learning for image recognition.
\newblock {\em Proc. IEEE Comput. Soc. Conf. Comput. Vis. Pattern Recognit.}, pages 770--778, December 2015.

\bibitem{Krizhevsky2012-wg}
Alex Krizhevsky, Ilya Sutskever, and Geoffrey~E Hinton.
\newblock {ImageNet} classification with deep convolutional neural networks.
\newblock {\em Adv. Neural Inf. Process. Syst.}, 25, 2012.

\bibitem{Heinlein2024-mg}
Lukas Heinlein, Roman~C Maron, Achim Hekler, Sarah Haggenm{\"u}ller, Christoph Wies, Jochen~S Utikal, Friedegund Meier, Sarah Hobelsberger, Frank~F Gellrich, Mildred Sergon, Axel Hauschild, Lars~E French, Lucie Heinzerling, Justin~G Schlager, Kamran Ghoreschi, Max Schlaak, Franz~J Hilke, Gabriela Poch, S{\"o}ren Korsing, Carola Berking, Markus~V Heppt, Michael Erdmann, Sebastian Haferkamp, Konstantin Drexler, Dirk Schadendorf, Wiebke Sondermann, Matthias Goebeler, Bastian Schilling, Eva Krieghoff-Henning, and Titus~J Brinker.
\newblock Clinical melanoma diagnosis with artificial intelligence: Insights from a prospective multicenter study.
\newblock January 2024.

\bibitem{Bashkatov2005-mg}
A~N Bashkatov, E~A Genina, V~I Kochubey, and V~V Tuchin.
\newblock Optical properties of human skin, subcutaneous and mucous tissues in the wavelength range from 400 to 2000 nm.
\newblock {\em J. Phys. D Appl. Phys.}, 38(15):2543, July 2005.

\bibitem{Arimoto2005-fo}
Hidenobu Arimoto, Mariko Egawa, and Yukio Yamada.
\newblock Depth profile of diffuse reflectance near-infrared spectroscopy for measurement of water content in skin.
\newblock {\em Skin Res. Technol.}, 11(1):27--35, February 2005.

\bibitem{Yokogawa2017-hi}
Sozo Yokogawa, Itaru Oshiyama, Harumi Ikeda, Yoshiki Ebiko, Tomoyuki Hirano, Suguru Saito, Takashi Oinoue, Yoshiya Hagimoto, and Hayato Iwamoto.
\newblock {IR} sensitivity enhancement of {CMOS} image sensor with diffractive light trapping pixels.
\newblock {\em Sci. Rep.}, 7(1):3832, June 2017.

\bibitem{Cooksey2015-ak}
Catherine~C Cooksey, Benjamin~K Tsai, and David~W Allen.
\newblock Spectral reflectance variability of skin and attributing factors.
\newblock In {\em Radar Sensor Technology {XIX}; and Active and Passive Signatures {VI}}, volume 9461, pages 510--517. SPIE, May 2015.

\bibitem{Kasuya2019-dy}
Akira Kasuya, Masahiro Aoshima, Kensuke Fukuchi, Takatoshi Shimauchi, Toshiharu Fujiyama, and Yoshiki Tokura.
\newblock An intuitive explanation of dermoscopic structures by digitally reconstructed pathological horizontal top-down view images.
\newblock {\em Sci. Rep.}, 9(1):19875, December 2019.

\bibitem{Gulia2012-ta}
Andrea Gulia and Cesare Massone.
\newblock Advances in dermoscopy for detecting melanocytic lesions.
\newblock {\em F1000 Med. Rep.}, 4:11, June 2012.

\end{thebibliography}

\newpage
\appendix
\counterwithin{figure}{section}
\section{Dermoscopic images}

\begin{figure*}[!ht]
\centering
\includegraphics[width=\textwidth]{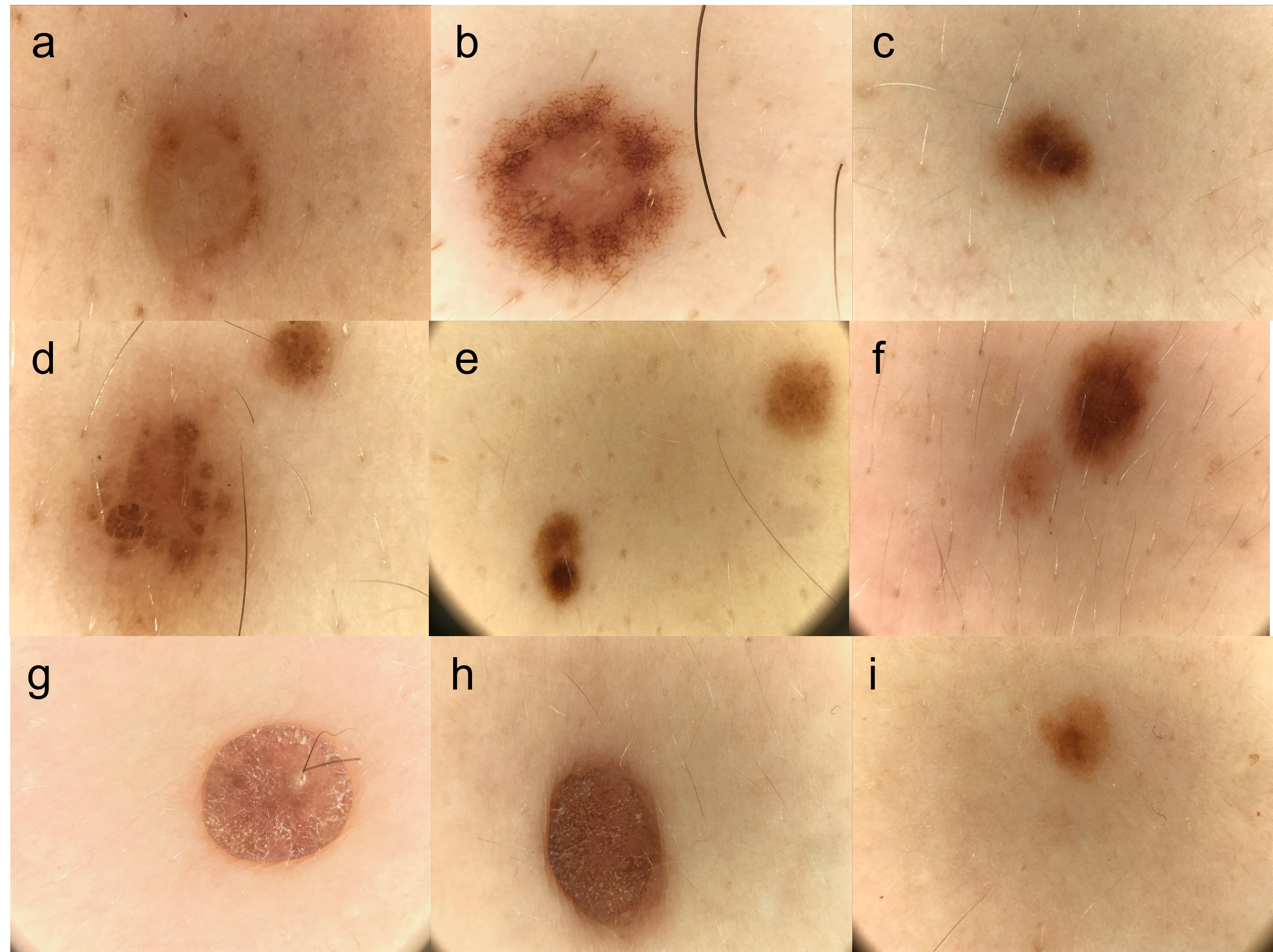}
\caption{
Dermoscopic images of different types of nevi of a single patient. The images a-i are labelled with their histological type and their dermoscopic pattern. a: compound nevus, 'fried egg' pattern; b: junctional melanocytic nevus, peripheral network with central hypopigmentantion; c: junctional melanocytic nevus, diffuse reticual pattern; d: compound nevus, globular pattern; e: junctional melanocytic nevus, diffuse reticual pattern; f: junctional melanocytic nevus, homogeneous brown; g: dermal nevus, homogeneous brown, h: dermal nevus, homogeneous brown, i: junctional melanocytic nevus, diffuse reticual pattern       
}
\label{fig:derma_images}
\end{figure*}

\end{document}